\documentstyle[12pt]{article}
\catcode`\@=11 \@addtoreset{equation}{section}\catcode`\@=12
\begin{document}
\begin{center}
{\large
EXACT SOLUTIONS IN MULTIDIMENSIONAL COSMOLOGY \\WITH SHEAR AND BULK 
VISCOSITY} \vskip 5mm 
V. R. Gavrilov, V.N. Melnikov\\

{\footnotesize {\it Center for Gravitation and Fundamental Metrology, 
     VNIIMS,\\ 3-1 Ulyanovoy St., Moscow, 117313, Russia}\\ 
     e-mail: melnikov@fund.phys.msu.su}\\
\vskip 3mm

     and\\
 
\vskip 3mm
 Roland Triay\\

{\footnotesize   {\it CNRS, Centre de Physique Theorique, Luminy,\\
Case 907,-F13288, Marseille, Cedex 9, France}\\
    e-mail: Roland.Triay@cpt.univ-mrs.fr}\\

\vspace{5mm}
{\bf Abstract}
\end{center}
Multidimensional cosmological model describing the evolution of a fluid 
with shear and bulk viscosity in $n$ Ricci-flat spaces is investigated. 
The barotropic equation of state for the density and the pressure in each 
space is assumed. 
The  second equation of state is chosen in the form when the bulk and the 
shear viscosity coefficients are inversely proportional to the volume of 
the Universe.
The 
integrability of Einstein equations reads as a colinearity constraint 
between vectors which are related to constant parameters in the first and 
second equations of state.  We give exact solutions in a Kasner-like form. 
The processes of dynamical compactification and the entropy production are 
discussed. The non-singular $D$-dimensional isotropic viscous solution is 
singled out.


\section{Introduction}
Last two decades have witnessed an increase of interest to the 
multidimensional cosmology (see, for instance, 
[5-7,9,11-18,20-25,27-29,31,32,39,41,42,45-48]
and references therein). \linebreak According to this theory
one assumes that the Universe had a higher dimension at a 
very early stage of its evolution, and that quantum processes have been 
responsible for the (topological) partition of the space, which provides 
us at present with the usual $3$-dimensional ({\em external\/}) space, in 
addition to {\em internal\/} space(s). The manifold which accounts for 
such a multidimensional spacetime has the following topology 
\begin{equation}\label{1.1}
 M = R \times M_{1} \times \ldots \times M_{n},
\end{equation}
where $R$ stands for the cosmic time axis, and the product with one part 
of manifolds $M_{1},\ldots,M_{n}$ gives the external space, when the 
remaining part stands for internal spaces. The classical stage of the 
evolution is governed by the multidimensional version of Einstein's 
equations. According to the classical description, the Riemann curvatures 
of spaces $M_1,\ldots,M_n$ are assumed to be constant (Einstein spaces). 
Such a model is the simplest multidimensional generalization of the 
space-time upon which the Friedman-Robertson-Walker (FRW) world model  is 
based on. One easily understands that the use of extra dimensions (for the 
physical space-time) can be a sensible scenario only at primordial 
epochs, since the standard FRW world model is known to be in a 
sufficiently good agreement with the observational constraints down to a 
quite primordial epoch such as the nucleosynthesis era. Hence, it is clear 
that a reduction process (called {\em dynamical  compactification\/} of 
additional dimensions) is required before such an epoch, to make the 
internal spaces contracting themself down to unobservable sizes.

Herein, we assume that the cosmic fluid (the source of the gravitational 
field at early stages) is viscous, which might simulate high energy 
physics processes (such as the particles creation).  The effects related 
to viscosity in $4$-dimensional Universe were studied through different 
viewpoints  (see e.g., [2-4,8,10,26,30,33-38,44]).
Before developing the 
\linebreak
multidimensional model let us briefly discuss
(extensive review of the subject was given by Gron \cite{Gron90}) the main
trends in $4$-dimensional cosmology with viscous fluid
as a source.

First, Misner \cite{Misner68} considered neutrino 
viscosity as a mechanism of reducing the anisotropy in the Early Universe. 
Stewart \cite{Stewart68} and Collins and Stewart 
\cite{CollinsStewart71} proved that it is 
possible only if initial anisotropies are small enough. Another series of 
papers which concerns the production of entropy in the viscous Universe 
was started by Weinberg \cite{Weinberg71}.  Both isotropization and 
production of entropy during lepton era in models of Bianchi types I,V 
were considered by Klimek \cite{Klimek76}. Caderni and Fabbri 
\cite{CaderniFabbri78}
calculated coefficients of shear and bulk viscosity in plasma and lepton 
eras within the model of Bianchi type I. The next approach is connected 
with obtaining singularity free viscous solutions. The first nonsingular 
solution was obtained by Murphy \cite{Murphy73} within the flat 
FRW model with fluid possessing a bulk viscosity. 
However, Belinsky and Khalatnikov 
\cite{BelinskyKhalatnikov75,BelinskyKhalatnikov76} showed that this 
solution corresponds to the very peculiar choice of parameters and is 
unstable with respect to the anisotropy  perturbations. Other 
nonsingular solutions with bulk viscosity were obtained by Novello and 
Araujo \cite{NovelloAraujo88}, Romero \cite{Romero88}, Oliveira and Salim 
\cite{OliveiraSalim88}.

The crucial feature of each viscous cosmological model is assuming of the
so called "second equation of state", which provides us with  
the viscosity coefficients dependence on time. Further we denote
by $\zeta$ and $\eta$ the bulk and shear viscosity coefficients,
correspondingly. 
Murphy \cite{Murphy73} integrated the 
4-dimensional flat FRW model with bulk viscosity by 
assuming $\zeta\sim\rho$ (as second equation of state), where $\rho$ is 
the density of the viscous fluid.  Belinsky and Khalatnikov 
\cite{BelinskyKhalatnikov75,BelinskyKhalatnikov76} studied the behavior of 
this model 
 as well as homogeneous anisotropic models of Bianchi types I and IX
by means of qualitative methods
with more general second 
equations of state $\zeta,\eta\sim\rho^{\nu}$, where $\nu$ is constant. 
Lukacs \cite{Lukacs76} integrated the homogeneous and isotropic 
$4$-dimensional model with a viscous pressureless fluid and a second equation of state given by
$\zeta\sim[\rm{scale\ factor}]^{-1}$. A curvature-dependent bulk viscosity was studied in
multidimensional cosmology by Wolf \cite{Wolf89}. Recently, Motta and Tomimura
\cite{MottaTomimura92} studied a $4$-dimensional inhomogeneous cosmology with a bulk viscosity
coefficient which depends on the metric. 
In our previous papers \cite{GavrilovEtal95a,GavrilovEtal96a} exact 
solutions in multidimensional models with bulk viscosity were obtained 
and their properties were studied for the following type of the second 
equation of state: $\zeta\sim[{\rm\ volume\ of\ the\ Universe}]^{-1}$.

The aim of the present investigation is to integrate the Einstein 
equations for a \linebreak
multidimensional cosmological model formed by a chain of 
Ricci-flat spaces and a cosmic fluid possessing both shear and bulk 
viscosity. The second equations of state are chosen in the following form 
of metrical dependence of the bulk and shear viscosity coefficients:
\linebreak
$\zeta,\eta\sim[{\rm volume\ of\ the\ Universe}]^{-1}$.

The paper is organized as follows.
In Sec. 2 we describe the model and get basic 
equations.  To integrate them, we develop some vector formalism proposed 
in our previous papers  \cite{GavrilovEtal95,GavrilovEtal96}.
Thermodynamical 
concepts in multidimensional cosmologies are defined in 
 Sec. 3, where a formula which provides us with the 
 variation rate of the entropy is derived. The equations of motion for the 
special set of parameters in the first and the second equations of state 
are integrated in Sec. 4. Exact solutions are given in a 
 Kasner-like form, their physical properties are investigated in 
 Sec. 5, where the process of dynamical compactification and 
the entropy production are also defined.
%%%%%%%%%%%%%%%%%%%%%%%%%%%%%%%%%%%%%%%%%%%%%%%%%%%%%%%%%%%%%%%%%%%%%%%%%
%%%%%%%%%%%%%%%%%%%%%%%%%%%%%%%%%%%%%%%%%%%%%%%%%%%%%%%%%%%%%%%%%%%%%%%%%
\section{The model}
Let us have the following metric
\begin{equation}\label{2.1}
{\rm d}s^2=-{\rm e}^{2\gamma(t)}{\rm d}t^2  + \sum_{i=1}^{n}\exp[2x^{i}(t)]{\rm d}s_i^2,
\end{equation}
 on the manifold defined in Eq.\,(\ref{1.1}), where ${\rm d}s_i^2$ is the 
 metric of the Einstein space $M_{i}$, $\gamma(t)$ and $x^{i}(t)$ are 
scalar functions of the cosmic time $t$. The dimension 
of this manifold is given by $D=1+\sum_{i=1}^{n}N_{i}$, where $N_{i}={\rm 
dim}\,M_{i}$. Herein, for reason of simplicity, only Ricci-flat spaces
$ M_{1},\ldots,M_{n}$ are assumed (i.e., the components of the Ricci 
tensor for the metrics ${\rm d}s_i^2$ are zero). One easily obtains the 
non-zero components of the Ricci-tensor for the metric defined in 
Eq.\,(\ref{2.1}) (see \cite{Ivashchuk92}):
\begin{eqnarray}
R_{0}^{0}
&=&{\rm e}^{-2\gamma(t)} \left(\sum_{i=1}^{n} N_{i}(\dot{x}^{i})^2 + \ddot{\gamma}_{0} -
\dot{\gamma}\dot{\gamma}_{0}\right),\label{2.2}\\
R_{n_{i}}^{m_{i}}
&=&{\rm e}^{-2\gamma(t)}\left(\ddot{x}^{i} + (\dot{\gamma}_{0} -
\dot{\gamma})\dot{x}^{i}\right)\delta_{n_{i}}^{m_{i}},\label{2.3}
\end{eqnarray}
for $i=1,\ldots,n$, where the indices $m_{i}$ and $n_{i}$ run from $D-\sum_{j=i}^{n}N_{j}$ to
$D-\sum_{j=i}^{n}N_{j}+N_{i}$ and $\gamma_{0}=\sum_{i=1}^{n}N_{i}x^{i}$.

A viscous fluid is characterized by a density $\rho$, a pressure $p$, a bulk viscosity
coefficient $\zeta$, a shear viscosity coefficient $\eta$, so that the (standard form of the)
energy-momentum tensor reads
\begin{equation}\label{2.5}
T_{\nu}^{\mu}=\rho u^{\mu}u_{\nu} + (p-\zeta \theta)P_{\nu}^{\mu}-2\eta\sigma_{\nu}^{\mu},
\end{equation}
where $u^{\mu}$ is the $D$-dimensional velocity of the fluid, $\theta=u^{\mu}_{;\mu}$ denotes the
scalar expansion, $P_{\nu}^{\mu}=\delta_{\nu}^{\mu} + u^{\mu}u_{\nu}$ is the projector on the
$(D-1)$-dimensional space orthogonal to $u^{\mu}$, and
$\sigma_{\nu}^{\mu}=
\frac{1}{2}\left(u_{\alpha;\beta}+u_{\beta;\alpha}\right)P^{\alpha\mu}P^{\beta}_{\nu}-
(D-1)^{-1}\theta P^{\mu}_{\nu}$ is the traceless shear tensor (defined as usual).

By choosing the $D$-dimensional velocity so that $u^{\mu}=\delta_{0}^{\mu}{\rm e}^{-\gamma(t)}$
 (the comoving observer condition), we obtain
\begin{eqnarray}
\theta&=&\dot{\gamma}_{0}{\rm e}^{-\gamma(t)},\label{2.9}\\
(u^{\mu}u_{\nu})&=&\rm diag(-1,0,\ldots,0),\label{2.10}\\
(P_{\nu}^{\mu})&=&\rm diag(0,1,\ldots,1),\label{2.11}\\
(\sigma^{\mu}_{\nu})&=&{\rm e}^{-\gamma}{\rm diag}\left(0,
\left(\dot{x}^1-\frac{\dot{\gamma}_0}{D-1}\right)\delta^{k_1}_{l_1},
\ldots,
\left(\dot{x}^n-\frac{\dot{\gamma}_0}{D-1}\right)\delta^{k_n}_{l_n}
\right),\label{2.12}
\end{eqnarray}
where $k_i$, $l_i=1,\ldots,N_i$ for $i=1,\ldots,n$.  
The function $\gamma(t)$ determines a time gauge (a harmonic 
time gauge for $\gamma(t)=\gamma_{0}$ and a synchronous time gauge for 
$\gamma(t)=0$), see Eq.\,(\ref{2.1}); note that the harmonic time $t$ and 
the synchronous time $t_s$ are related by ${\rm d}t_s=\exp[\gamma_{0}]{\rm 
d}t$.  By assuming anisotropy properties for the pressure and the bulk 
viscosity, with respect to the whole space $M_{1}\times\ldots\times 
M_{n}$, one has
 \begin{equation}\label{2.13}
(T^{\mu}_{\nu})={\rm diag}(-\rho,
p^*_1\delta^{k_1}_{l_1},
\ldots,
p^*_n\delta^{k_n}_{l_n}),
\end{equation}
where
\begin{equation}
p^*_i=p_i - {\rm e}^{-\gamma}\left[\zeta_i\dot{\gamma}_0+
2\eta\left(\dot{x}^i-\frac{\dot{\gamma}_0}{D-1}\right)\right],
\end{equation}
and $p_{i}$, resp. $\zeta_{i}$, is the pressure, resp. the bulk viscosity coefficient, in the
space described by the manifold $M_{i}$. Furthermore, we assume that the barotropic equations
of state hold
\begin{equation}\label{2.15}
p_{i} = (1-h_{i})\rho(t),
\end{equation}
where the $h_{i}$ are constants. One easily shows that the form of the equation of motion
($\bigtriangledown _{M} T_{0}^{M}=0$) for a viscous fluid described by a tensor given
by Eq.\,(\ref{2.13}), is given by
\begin{equation}\label{2.16}
\dot{\rho}+\sum_{i=1}^{n}N_{i}\dot{x}^{i}(\rho+p^*_{i})=0.
\end{equation}

The Einstein equations $R^{\mu}_{\nu}-\frac{1}{2}\delta^{\mu}_{\nu}R=\kappa^{2}T^{\mu}_{\nu}$, where
$\kappa^{2}$ is the gravitational constant, can be written as
$R^{\mu}_{\nu}=\kappa^{2}(T^{\mu}_{\nu}-\frac{T}{D-2}\delta^{\mu}_{\nu})$.
Further, by using the equations
$R^{0}_{0}-\frac{1}{2}\delta^{0}_{0}R=\kappa^{2}T^{0}_{0}$
and 
$R^{m_{i}}_{n_{i}}=\kappa^{2}(T^{m_{i}}_{n_{i}}- \frac{T}{D-2}\delta^{m_{i}}_{k_{i}})$,
Eqs.(\ref{2.2},\ref{2.3},\ref{2.13}) give the following equations of motion
\begin{equation} \label{2.17}
\sum_{i=1}^{n}N_{i}(\dot{x}^{i})^{2} - \dot{\gamma}_{0}^{2}=
-2\kappa^{2}{\rm e}^{2\gamma}\rho,
\end{equation}
\begin{eqnarray}
\ddot{x}^{i} + (\dot{\gamma}_{0}-\dot{\gamma})\dot{x}^{i}&=&
\kappa^{2}{\rm e}^{\gamma}
\left[{\rm e}^{\gamma}\rho
\left(-h_{i} + \frac{\sum_{k=1}^{n}N_{k}h_{k}}{D-2}\right) \right.\nonumber\\
&+& \left.\dot{\gamma}_{0}
\left(-\zeta_{i} +\frac{\sum_{k=1}^{n}N_{k}\zeta_{k}}{D-2}\right)
\right]-2\kappa^{2}{\rm e}^{\gamma}\eta
\left(\dot{x}^i-\frac{\dot{\gamma}_0}{D-1}\right).\label{2.18}
\end{eqnarray}

We use an integration procedure which is based on the $n$-dimensional real vector space
$R^{n}$. Let $e_{1},\ldots,e_{n}$ be the canonical basis in $R^{n}$ (i.e. 
$e_{1}=(1,0,\ldots,0)$ etc\ldots), and $\langle,\rangle$ denote a 
symmetrical bilinear form defined on $R^{n}$ by  
\begin{equation} 
\langle 
e_{i},e_{j}\rangle=\delta_{ij}N_{j} - N_{i}N_{j}\equiv G_{ij}.  
\end{equation}
It has been used as a mini-super-space metric for cosmological models (see 
[20-25]).
Such a form is non-generate and has the pseudo-Euclidean signature  
$(-,+,\ldots,+)$. 
With
this in mind, a vector $y\in R^{n}$ is time-like, resp. space-like or isotropic, if $\langle
y,y\rangle$ takes negative, resp. positive or null values; and two vectors $y$ and $z$ are
orthogonal if $\langle y,z\rangle=0$. Hereafter, we use the following vectors
\begin{eqnarray}
x&=&x^{1}e_{1}+\ldots+x^{n}e_{n},\label{2.23}\\
u&=&u^{1}e_{1}+\ldots+u^{n}e_{n},\quad
u^{i}=h_{i}-\frac{\sum_{k=1}^{n}N_{k}h_{k}}{D-2},\quad u_{i}=N_{i}h_{i},\label{2.24}\\
\xi&=&\xi^{1}e_{1}+\ldots+\xi^{n}e_{n},\quad
\xi^{i}=\zeta_{i}-\frac{\sum_{k=1}^{n}N_{k}\zeta_{k}}{D-2},\quad
\xi_{i}=N_{i}\zeta_{i}\label{2.25},
\end{eqnarray}
where covariant coordinates of the vectors are introduced by the usual 
way.  Moreover, let us denote $u_{d}$ the particular vector given by 
Eq.\,(2.17) with $h_{i=1,\ldots,n}=1$ (it is related to dust in the 
whole space, see Eq.\,(\ref{2.15})). One has
\begin{equation}
(u_{d})_{i}=N_{i},\quad u_{d}^{i}=\frac{-1}{D-2},\quad
\langle u_{d},u_{d}\rangle=-\frac{D-1}{D-2},\quad
\langle u_{d},x\rangle=\gamma_{0}.
\end{equation}
Thus, using Eqs.\,(2.16-2.19) we rewrite the Einstein equations 
(2.13),(2.14) in the form
\begin{eqnarray}
\langle \dot{x},\dot{x}\rangle&=&-2\kappa^{2}{\rm e}^{2\gamma}\rho,\label{2.27}\\
\ddot{x} + \left(\langle u_{d},\dot{x}\rangle-\dot{\gamma}+
2\eta\kappa^2{\rm e}^{\gamma}\right)\dot{x}&=&
\frac{\langle \dot{x},\dot{x}\rangle}{2}u \nonumber\\
&-&\kappa^{2}{\rm e}^{\gamma}\langle u_{d},\dot{x}\rangle\left(
\xi-\frac{2\eta}{\langle u_d,u_d\rangle}u_d\right),\label{2.30}
\end{eqnarray}
where the formal dependence on $\rho$ in Eq.\,(2.21) has been canceled, 
according to Eq.\,(2.20). Moreover, Eq.\,(\ref{2.16}) can be written 
as 
\begin{equation} 
\dot{\rho}+\langle 2u_{d}-u,\dot{x}\rangle\rho- {\rm 
e}^{-\gamma} \left(2\eta\langle \dot{x},\dot{x}\rangle+ \langle 
u_{d},\dot{x}\rangle\langle \xi-\frac{2\eta u_d}{\langle 
u_d,u_d\rangle},\dot{x}\rangle\right)=0.  
\end{equation} 
To integrate 
Eq.(2.21) one needs second equations of state, involving the bulk 
viscosity coefficients $\zeta_1,\ldots,\zeta_n$ and the shear viscosity 
coefficient $\eta$. Let us assume that these coefficients are proportional 
to $\exp[-\gamma_0]$ (or inversely proportional to the volume of the 
Universe), i.e.  \begin{equation}\label{2.31} \eta,\zeta_{i} \sim [{\rm 
scale~factor~of}\,M_{1}]^{-\rm{dim}(M_{1})} \cdot\ldots\cdot [{\rm 
scale~factor~of}\,M_{n}]^{-\rm{dim}(M_{n})}, \end{equation} which means 
 (from a physical viewpoint) that the viscosity decreases when the space 
 $M_{1}\times \ldots\times M_{n}$ expands. The integrability of the basic 
equation (provided the second equations of state) is ensured when the 
vectors $u,\xi,u_d$ are either colinear or orthogonal (with respect to the 
mini-super-space metric) in some combination [15,18].  
Herein, we suppose that these vectors are colinear, which means that the 
viscous fluid has identical properties in the internal space(s) and the 
external space. Hence, all these assumptions, for the pressures and the 
 viscosity coefficients, allow us to write \begin{eqnarray} 
p_{i}&=&(1-h)\rho \quad {\rm or} \quad u=hu_{d},\label{2.32}\\ 
\zeta_{i}&=&\frac{\zeta_{0}}{\kappa^{2}}{\rm e}^{-\gamma_{0}}\equiv\zeta 
\quad {\rm or}            \quad \xi=\zeta u_d=\frac{\zeta_{0}} 
{\kappa^{2}}{\rm e}^{-\gamma_{0}}u_{d},\quad (i=1,\ldots,n),\label{2.33}\\ 
\eta &=&\frac{\eta_{0}}{2\kappa^{2}}{\rm e}^{-\gamma_{0}},\label{2.34} 
\end{eqnarray} where $\zeta_{0},\eta_0$ and $h$ are constants.  
%%%%%%%%%%%%%%%%%%%%%%%%%%%%%%%%%%%%%%%%%%%%%%%%%%%%%%%%%%%%%%%%%%%%%%%%%%%%%
%%%%%%%%%%%%%%%%%%%%%%%%%%%%%%%%%%%%%%%%%%%%%%%%%%%%%%%%%%%%%%%%%%%%%%%%%%%%%%%
%%%%%%%%%%%%%%%%%%%%%%%%%%%%%%%%%%%%%%%%%%%%%%%%%%%%%%%%%%%%%%%%%%%%%%%%%%%%%

\section{Multidimensional Thermodynamics}
 According to [43,48], let us summarize thermodynamics 
principles in such a \linebreak
multidimensional cosmology. The first law of 
thermodynamics reads 
\begin{equation}\label{3.1} T{\rm d}S={\rm d}(\rho V) 
+ V\sum_{i=1}^{n}p_{i}\frac {{\rm d}V_{i}}{V_{i}}, \end{equation} where 
$V_{i}$ stands for a fluid volume in the space $M_{i}$, when $V=V_{1}\cdot 
\ldots \cdot V_{n}$ is a fluid volume in the whole space, and $S$ is an 
entropy in the volume $V$. By assuming that the baryon particle number 
$N_{\rm B}$ in the volume $V$ is conserved, Eq.\,(\ref{3.1}) transforms to
\begin{equation}\label{3.2}
nT\dot{s}=\dot{\rho}+\rho\sum_{i=1}^{n}N_{i}\dot{x}^{i}+
 \sum_{i=1}^{n}p_{i}N_{i}\dot{x}^{i},
\end{equation}
where $s=S/N_{\rm B}$, resp. $n=N_{\rm B}/V$, stands for the entropy per baryon, resp. the baryon
number density. Let us remind that ${\exp[x^{i}]}$ is the scale factor of the space $M_{i}$
(of dimension $N_{i}$). For a perfect fluid ($\zeta_{i}=0$, $\eta=0$), the comparison between
Eqs.\,(\ref{2.16},\ref{3.2}) gives the entropy conservation ( i.e., $s$ is 
constant). 
Similarly, we obtain also the temperature (see \cite{Zhuk95}).  From 
Eq.\,(\ref{3.2}) we have
\begin{equation}\label{3.4}
\left(\frac{\partial\rho}{\partial x^{i}}\right)_{s,x^{j}}=-\rho
N_{i}-p_{i}N_{i}=(h_{i}-2)N_{i}\rho,\quad (j\neq i),
\end{equation}
and then
\begin{equation}\label{3.5}
\rho=K(s)\exp\left[\sum_{i=1}^{n}(h_{i}-2)N_{i}x^{i}\right],
\end{equation}
where $K(s)$ is an unknown function (which reads in term of the  entropy 
per baryon $s$). Using Eqs.\,(3.2),(3.4) we get
\begin{equation}\label{3.6}
\left(\frac{\partial\rho}{\partial s}\right)_{x^{i}}=nT=K^{\prime}(s)
\exp\left[\sum_{i=1}^{n}(2-h_{i})N_{i}x^{i}\right].
\end{equation}
For a perfect fluid, we have $K^{\prime}(s)=1/B$
where $B$ is a constant, then
\begin{equation}\label{3.8}
nT=\frac{1}{B}\exp\left[\sum_{i=1}^{n}(h_{i}-2)N_{i}x^{i}\right]=
\frac{1}{B}\exp[\langle u-2u_{d},x\rangle].
\end{equation}

 For a fluid with a bulk and shear viscosity, the comparison between  
 Eqs.\,(\ref{2.16},\ref{3.2}) provides us with 
\begin{equation}\label{3.9} 
nT\dot{s}= {\rm e}^{-\gamma}\left(2\eta\langle \dot{x},\dot{x}\rangle
+\langle u_d,\dot{x}\rangle\langle \xi-2\frac{2\eta}{\langle u_d,u_d\rangle}u_d,\dot{x}\rangle\right).
\end{equation}
Such a formula gives the variation rate of entropy per baryon  in 
multidimensional cosmology on the manifold $M = R \times M_{1} \times 
\ldots \times M_{n}$ with viscosity.  The entropy production can be 
calculated if the temperature of the fluid is known. Herein, we suppose 
that the temperature is given by the perfect fluid formula 
Eq.\,(3.6), which is valid with sufficient accuracy when effects 
of viscosity  are small. Hence, Eqs.\,(3.6),(3.7) give 
\begin{equation}\label{3.10}
\dot{s}=
B{\rm e}^{\langle 2u_d-u,x\rangle-\gamma}
\left(2\eta\langle \dot{x},\dot{x}\rangle
+\langle u_d,\dot{x}\rangle\langle \xi-2\frac{2\eta u_d}{\langle 
u_d,u_d\rangle},\dot{x}\rangle\right).  \end{equation}

%%%%%%%%%%%%%%%%%%%%%%%%%%%%%%%%%%%%%%%%%%%%%%%%%%%%%%%%%%%%%%%%%%%%%%%%%
%%%%%%%%%%%%%%%%%%%%%%%%%%%%%%%%%%%%%%%%%%%%%%%%%%%%%%%%%%%%%%%%%%%%%%%%%%
%%%%%%%%%%%%%%%%%%%%%%%%%%%%%%%%%%%%%%%%%%%%%%%%%%%%%%%%%%%%%%%%%%%%%%%%%%
%%%%%%%%%%%%%%%%%%%%%%%%%%%%%%%%%%%%%%%%%%%%%%%%%%%%%%%%%%%%%%%%%%%%%%%%%%
\section{Exact solutions}

According to assumptions given in 
Eqs.\,(2.24-2.26), the basic vector equation, see  
Eq.\,(2.21), reads
\begin{eqnarray}\label{4.1} 
\ddot{x}+\left(\langle u_d,\dot{x}\rangle-\dot{\gamma}+2\eta\kappa^2{\rm
e}^{\gamma}\right)\dot{x}=
\left[\frac{h}{2}\langle \dot{x},\dot{x}\rangle-\kappa^2{\rm e}^{\gamma}\langle u_d,\dot{x}\rangle
\left(\zeta-\frac{2\eta}{\langle u_d,u_d\rangle}\right)\right]u_d.
\end{eqnarray}
In order to integrate such a (vector) equation, we use the orthogonal basis
\begin{equation}\label{4.2}
\frac{u_d}{\langle u_d,u_d\rangle},f_2,\ldots,f_n\in R^n,
\end{equation}
where the orthogonality property reads
\begin{equation}\label{4.3}
\langle u_d,f_j\rangle=0,\ \ \langle f_j,f_k\rangle=\delta_{jk},\quad (j,k=2,\ldots,n).
\end{equation}
Let us note that the basis vectors $f_2,\ldots,f_n$ are space-like, since they are orthogonal to
 the time-like vector $u_d$. The vector $x\in R^n$ decomposes as follows
\begin{equation}\label{4.4}
x=\langle u_d,x\rangle\frac{u_d}{\langle u_d,u_d\rangle}+\sum_{j=2}^n\langle f_j,x\rangle f_j.
\end{equation}
Hence, the basic vector equation given in Eq.\,(\ref{4.1}) reads in term of coordinates in such a
basis as follows
\begin{equation} \label{4.5}
\langle f_j,\ddot{x}\rangle+\left(\langle u_d,\dot{x}\rangle-\dot{\gamma}+2\eta\kappa^2{\rm
e}^{\gamma}\right)\langle f_j,\dot{x}\rangle = 0
\end{equation}
\begin{equation}\label{4.6}
\frac{\langle u_d,\ddot{x}\rangle}{\langle 
u_d,u_d\rangle}+\left(\frac{\langle u_d,\dot{x}\rangle-\dot{\gamma}} 
{\langle u_d,u_d \rangle}+ \zeta\kappa^2{\rm e}^{\gamma}\right)\langle 
u_d,\dot{x}\rangle = \frac{h}{2}\left[\frac{\langle 
u_d,\dot{x}\rangle^2}{\langle u_d,u_d\rangle} +\sum_{j=2}^n\langle 
f_j,\dot{x}\rangle^2\right], 
\end{equation} 
which provides us with a set of 
equations (for $j=2,\ldots,n$). For the metric dependence of the viscosity 
coefficients, see Eqs.\,(2.25),(2.26) 
Eqs.\,(\ref{4.5},\ref{4.6}) read 
\begin{equation}\label{4.7} \langle 
f_j,\ddot{x}\rangle+\eta_0\langle f_j,\dot{x}\rangle=0,\quad 
(j=2,\ldots,n) \end{equation} \begin{equation}\label{4.8} \langle 
u_d,\ddot{x}\rangle-\frac{h}{2}\langle u_d,\dot{x}\rangle^2 +\langle 
u_d,u_d\rangle\left(\zeta_0\langle 
u_d,\dot{x}\rangle-\frac{h}{2}\sum_{j=2}^n\langle 
f_j,\dot{x}\rangle^2\right) =0
\end{equation} 
in the harmonic time gauge
\begin{equation}\label{4.9}
\gamma=\gamma_0=\langle u_d,x\rangle=\sum_{i=1}^n N_ix^i.
\end{equation}
Such a set of equations is integrable for any values of the constant parameters $h$, $\eta_0$ and
$\zeta_0$.

Let us first assume models with
\begin{equation}\label{4.10}
h\neq 0.
\end{equation}
The integration of Eq.\,(\ref{4.7}) gives
\begin{equation}\label{4.11}
\langle f_j,x\rangle=\left\{
\begin{tabular}{ll}
$tp^j+q^j$ & if $\eta_0=0$,\\
${\rm e}^{-\eta_0t}p^j+q^j$ & if $\eta_0\neq 0$,
\end{tabular}\right.
\end{equation}
where $p^j$ and $q^j$ are arbitrary constants. The Kasner-like form solution can be written in
term of vectors $\alpha,\beta\in R^n$, defined as follows
\begin{equation}\label{4.13}
\alpha=\sum_{j=2}^n p^jf_j\equiv\sum_{i=1}^n\alpha^ie_i,\quad
\beta=\sum_{j=2}^n q^jf_j\equiv\sum_{i=1}^n\beta^ie_i,
\end{equation}
where $\alpha^i$ and $\beta^i$ are their coordinates in the canonical basis $e_1,\ldots,e_n$. By
using the orthogonality conditions, we obtain
\begin{equation}\label{4.14}
\langle \alpha,u_d\rangle=\sum_{i=1}^n\alpha^iN_i=0,\quad
\langle \beta,u_d\rangle=\sum_{i=1}^n\beta^i N_i=0,
\end{equation}
\begin{equation}\label{4.15a}
\langle \alpha,\alpha\rangle
=\sum_{i=1}^n \left(\alpha^i\right)^2N_i
= \sum_{j=2}^n\left(p^j\right)^2\geq 0
\end{equation}
\begin{equation}\label{4.15b}
\langle \beta,\beta\rangle
=\sum_{i=1}^n \left(\beta^i\right)^2N_i
=\sum_{j=2}^n\left(q^j\right)^2\geq0,
\end{equation} where the constants $\alpha^i$ and $\beta^i$ may be called Kasner-like parameters,
because of the existence of these constraints. By using 
Eqs.\,(\ref{4.4},\ref{4.11},\ref{4.13}), we obtain 
\begin{equation}\label{4.16}
x=\langle u_d,x\rangle\frac{u_d}{\langle u_d,u_d\rangle}+a(t)\alpha+\beta,
\end{equation}
where the function
\begin{equation}\label{4.17}
a(t)=
\left\{
\begin{tabular}{ll}
$t$ & if $\eta_0=0$,\\
${\rm e}^{-\eta_0t}$ & if $\eta_0\neq 0$,
\end{tabular}\right.
\end{equation}
By substituting the functions $\langle f_j,x\rangle$ into Eq.\,(\ref{4.8}) we obtain the
following equation for the unknown function $\langle u_d,x\rangle$
\begin{equation}\label{4.19}
\langle u_d,\ddot{x}\rangle-\frac{h}{2}\langle u_d,\dot{x}\rangle^2+
\langle u_d,u_d\rangle\zeta_0\langle u_d,\dot{x}\rangle=
-\frac{h}{2}A^2\dot{a}^2(t),
\end{equation}
where
\begin{equation}\label{4.20}
A=\sqrt{\frac{D-1}{D-2}\langle \alpha,\alpha\rangle}.
\end{equation}
The equation Eq.\,(\ref{4.19}) has been integrated for the non-viscous model $\eta_0=\zeta_0=0$
(see, e.g., \cite{GavrilovEtal95}) and for the model with $\zeta_0\neq 0$ and $\eta_0=0$ in
\cite{GavrilovEtal95a,GavrilovEtal96a}. For the model with $\eta_0\neq 0$, 
it can be reduced to the modified Bessel equation 
\begin{equation}\label{4.21}
\tau^2\frac{{\rm d}^2z}{{\rm d}\tau^2}+\tau\frac{{\rm d}z}{{\rm d}\tau}-(\tau^2+\nu^2)z=0
\end{equation}
by means of the transformation
\begin{equation}\label{4.22}
\tau=\frac{1}{2}|h|A{\rm e}^{-\eta_0t},
\end{equation}
\begin{equation}\label{4.23}
\langle u_d,\dot{x}\rangle=\frac{2\eta_0}{h}\tau
\frac{{\rm d}}{{\rm d}\tau} \ln|\tau^{-\nu} z(\tau)|
\end{equation}
in the non-trivial case $\langle\alpha,\alpha\rangle\neq 0$, where the constant
\begin{equation}\label{4.25}
\nu=\frac{D-1}{D-2}\frac{\zeta_0}{2\eta_0}.
\end{equation}
The general solution of the modified Bessel equation is given by
\begin{equation}\label{4.26}
z(\tau)=C_1I_{|\nu|}(\tau)+C_2K_{|\nu|}(\tau),
\end{equation}
see \cite{Handbook}, where $I_{|\nu|}(\tau)$, resp. $K_{|\nu|}(\tau)$, is the related modified
Bessel function, resp. Mac-Donald function.

Finally, the results of Eqs.\,(\ref{4.7},\ref{4.8}) integration 
for various constants $\zeta_0$ and $\eta_0$ can be 
presented in Tab.\,1

\begin{center}
Table 1\\
\vskip 5mm
\begin{tabular}{|l|p{6cm}|p{6cm}|}
\hline
\ & \ & \ \\
\ & $\zeta_0=0$ & $\zeta_0\neq 0$\\
\ & \ & \ \\
\hline
\ & \ & \ \\
$\eta_0=0$ & Solution II with $\langle \alpha,\alpha\rangle\neq 0$ and 
Solution I & Solution II 
\\ \ & \ & \ \\ \hline \ & \ & \ \\ $\eta_0\neq 
0$ & Solution III with $\langle \alpha,\alpha\rangle\neq 0$ and Solution I 
& Solution III with $\langle \alpha,\alpha\rangle\neq 0$ and Solution II 
with $\langle \alpha,\alpha\rangle= 0$
\\ \ & \ & \ 
\\ \hline 
\end{tabular} 
\end{center}

%\end{flushleft}
%\end{table}
\vskip 5mm

The solutions I,II,III in term of scale factors are the following

\begin{itemize} \item Solution~I~:
\begin{eqnarray}
{\rm e}^{x^i}&=&{\rm e}^{\beta^i}\left|C_1+C_2t\right|^{-2/[h(D-1)]}.
\end{eqnarray}
\item Solution~II~:
\begin{eqnarray}
{\rm e}^{x^i}
&=&{\rm e}^{\alpha^it+\beta^i}
\left|
C_1{\rm e}^{-(\tilde{A}-\tilde{\zeta}_0)ht/2}+ C_2{\rm e}^{(\tilde{A}+\tilde{\zeta}_0)ht/2}
\right|^{-2/[h(D-1)]},
\end{eqnarray}
where
\begin{equation}
\tilde{\zeta}_0=\frac{D-1}{D-2}\frac{\zeta_0}{h},\quad
\tilde{A}=\sqrt{\frac{D-1}{D-2}\langle \alpha,\alpha\rangle+
\left(\frac{D-1}{D-2}\frac{\zeta_0}{h}\right)^2}=
\sqrt{A^2+\tilde{\zeta}_0^2}.
\end{equation}
\item Solution~III~: 
\begin{eqnarray}
{\rm e}^{x^i}&=&\exp{\left(\alpha^i{\rm e}^{-\eta_0t}+\beta^i\right)}
\left(
\tau^{-\nu}\left| C_1I_{|\nu|}(\tau)+C_2K_{|\nu|}(\tau) \right|
\right)^{-2/[h(D-1)]},
\end{eqnarray}
where the variable $\tau>0$ is given in Eq.\,(\ref{4.22}), the constants $A$ in Eq.\,(\ref{4.20})
and $\nu$ in Eq.\,(\ref{4.25}).
\end{itemize}
In formulas (4.25),(4.27),(4.28) $C_{i=1,2}$ are integration constants 
such that $C_1^2+C_2^2>0$ and $i=1,\ldots,n$. The Kasner-like parameters
$\alpha^i$ and $\beta^i$ obey the relations given in Eq.\,(4.13).

The set of equations given in Eqs.\,(\ref{4.7},\ref{4.8}) is easily 
integrable in the case 
\begin{equation}\label{4.42} h=0, 
\end{equation} 
which, with the barotropic equation of state in mind, relates to Zeldovich or stiff matter. The
results are given as follows~:
\begin{itemize}
\item
Solution~IV~: for $i=1,\ldots,n$
\begin{eqnarray}
{\rm e}^{x^i}&=&\exp\left[\alpha^i
a(t)+\beta^i\right]\times\left\{
\begin{tabular}{ll}
$\exp\left(C_1+C_2t\right)$ &if $\zeta_0=0$\\
$\exp\left(C_1+C_2\exp\left(\zeta_0\frac{D-1}{D-2}t\right)\right)$ &if 
$\zeta_0\neq 0$, 
\end{tabular}\right.
\end{eqnarray} 
where $C_{i=1,2}$ are 
arbitrary constants, and the function $a(t)$ is given in 
Eq.\,(\ref{4.17}).  
\end{itemize} 
%%%%%%%%%%%%%%%%%%%%%%%%%%%%%%%%%%%%%%%%%%%%%%%%%%%%%%%%%%%%%%%%%%%
%%%%%%%%%%%%%%%%%%%%%%%%%%%%%%%%%%%%%%%%%%%%%%%%%%%%%%%%%%%%%%%%%%%%%
%%%%%%%%%%%%%%%%%%%%%%%%%%%%%%%%%%%%%%%%%%%%%%%%%%%%%%%%%%%%%%%%%%%%%%
%%%%%%%%%%%%%%%%%%%%%%%%%%%%%%%%%%%%%%%%%%%%%%%%%%%%%%%%%%%%%%%%%%%%%%
%%%%%%%%%%%%%%%%%%%%%%%%%%%%%%%%%%%%%%%%%%%%%%%%%%%%%%%%%%%%%%%%%%%%
\section{Discussion}
 Let us remind the multidimensional generalization of the well-known 
{\em Kasner solution\/} \cite{IvashchukMelnikov94}, it reads (for the 
synchronous time $t_s$) as follows 
\begin{equation}\label{5.1} {\rm 
d}s^2=-{\rm d}t^2_s+\sum_{i=1}^{n}A_it_s^{2\varepsilon^i}{\rm d}s^2_i.  
\end{equation} 
Such a metric describes the evolution of a vacuum model defined on the manifold
$R\times M_{1}\times\ldots\times M_{n}$, where the $M_{i}$ are 
Ricci-flat spaces of dimension $N_i$ with the metric ${\rm d}s_i^2$, $A_i$ 
are arbitrary constants and $\varepsilon_i$ are the Kasner parameters, which 
satisfy the relations 
\begin{equation}\label{5.2} 
\sum_{i=1}^{n}N_i\varepsilon^i=1,\quad \sum_{i=1}^{n}N_i(\varepsilon^i)^2=1.
\end{equation}
The $\varepsilon^i$ and the Kasner-like parameters $\alpha^i$ (used in the above formulas for the
exact solutions) are related as
\begin{equation}\label{5.3}
\varepsilon^i=\pm\frac{\alpha^i}{A}+\frac{1}{D-1},\quad \langle 
\alpha,\alpha\rangle\neq 0.  
\end{equation} 

By using Eq.\,(\ref{4.16}) 
(i.e., a general decomposition of the vector $x\in R^n$) and the result of 
Sec.4, see Eq.\,(3.8), we obtain the variation rate 
of entropy 
\begin{equation}\label{5.4} 
\dot{s}=\frac{B}{\kappa^2}{\rm 
e}^{-h\gamma_0}\left(\zeta_0\dot{\gamma}_0^2+ \eta_0\langle 
\alpha,\alpha\rangle\dot{a}^2(t)\right), 
\end{equation} which shows that 
the entropy increases when $\zeta_0dt>0$ and $\eta_0dt>0$. Further we 
assume 
\begin{equation}\label{5.5} 
\zeta_0\geq 0,\quad \eta_0\geq 0,
\end{equation}
so harmonic time $t$ (as well as synchronous time $t_s$) increases
during the evolution.

One easily shows that the {\em weak energy condition\/} 
$T^{\mu}_{\nu}v^{\nu}v_{\mu}\geq0$, for any D-dimensional non space-like 
vector $v^{\nu}$ (and thus as well as for the 4-dimensional case), applied 
to the stress-energy tensor given in Eq.\,(\ref{2.13}), can be written as 
inequalities 
\begin{equation}\label{5.6} 
\rho\geq 0,\quad \rho+p^*_i\geq 
0,\quad (i=1,\ldots,n), 
\end{equation} where $\rho$, resp. $p_i^*$, is the 
density, resp. the effective pressure, of the fluid. 

The {\em dominant energy 
condition\/} $T^{\mu}_{\nu}v^{\nu}v_{\mu}\geq0$ and 
$T^{\nu}_{\mu}T^{\mu}_{\lambda}v_{\nu}v^{\lambda}\leq 0$, for any non 
space-like vector $v^{\mu}$, applied to the stress-energy tensor given in 
Eq.\,(\ref{2.13}), reduces to inequalities defined in Eq.\,(\ref{5.6}), 
with the following additional condition
\begin{equation}\label{5.7}
\rho-p_i^*\geq 0,\quad (i=1,\ldots,n)
\end{equation}
Notice that due to the weak energy condition 
the following restriction on the constant $h$ (taken from the barotropic 
equation of state (2.11)) arises in the nonviscous case: $h\leq 2$. The 
dominant energy condition for the nonviscous stress-energy tensor implies: 
$0\leq h\leq 2$.

It is important to note that Solution~I and
Solutions~II,IV for zero Kasner-like parameters ($\alpha^i=0,\quad i=1,\ldots,n$) are isotropic,
since the spaces $M_1,\ldots,M_n$ have identical scale factors.

Further we discuss the solutions obtained, which are of interest within 
multidimensional or $4$-dimensional cosmology.

\subsection{Non Viscous Models} 
For a better understanding of the 
viscosity effect on the dynamics, we first outline the properties of non 
viscous models.  

\subsubsection{}
 The {\em isotropic non viscous model\/} is 
described by  Solution~I ($h\neq 0$) and Solution~IV ($h=0$) with 
$\alpha^i=0$ (for $i=1,\ldots,n$), which represents the multidimensional 
generalization of the flat FRW model. It is the steady-state model 
\begin{equation}\label{5.8} {\rm e}^{x^i}\sim 
\exp\left[-\frac{C_2}{D-1}t_s \right], \ \ \rho={\rm const} 
\end{equation} 
for $h=2$ and shows a power-law behavior
\begin{equation}\label{5.9}
{\rm e}^{x^i}\sim t_s^{2/[(2-h)(D-1)]},\ \ \rho\sim t_s^{-2}
\end{equation}
for $h\neq 2$, where $t_s$ is the synchronous time (stationary solution 
with zero density is also possible); let us call it as Friedman-like 
behavior.  
 
\subsubsection{} 
The {\em anisotropic non 
viscous model\/} for $h\in (0,2]$ is described by Solution~II with 
$\zeta_0=0$ and $\langle \alpha,\alpha\rangle\neq 0$. One easily shows 
that the integration constants $C_{i=1,2}$ have to satisfy the condition 
$C_1C_2<0$, otherwise $\rho\leq0,\quad\forall t$. By using this condition, 
we obtain from Eq.\,(4.26)

\begin{equation}\label{5.10}
{\rm e}^{\gamma_0}\sim \left|\sinh[Ah(t-t_0)/2]\right|^{-2/h}
\end{equation}
where $t_0=0$ can be chosen (with no loss of generality). Then, for a  suitable choice of the integration constant of equation ${\rm 
d}t_s=\exp(\gamma_0){\rm d}t$, one has the following correspondences 
$t\in(-\infty,0)\Leftrightarrow t_s\in(0,+\infty)$ and 
$t\in(0,+\infty)\Leftrightarrow t_s\in(-\infty,0)$. Hence, we solely 
investigate the solution $t_s\in(-\infty,0)$, since the evolution of the 
non-viscous fluid is  reversible. 
 
 From Eq.\,(\ref{5.10}), we obtain in the main order 
$\exp[\gamma_0]\sim t^{-2/h}$ when $t\rightarrow +0$ ($t_s\rightarrow 
+\infty$), then one has $t_s\sim t^{1-2/h}$ for $h\in (0,2)$ and 
$t_s\sim\ln t$ for $h=2$. By using these relations and Eq.\,(4.26), 
we easily see that the multidimensional Universe shows an isotropical 
Friedman-like contraction, as defined by Eq.\,(\ref{5.9}), in the 
(infinite) past. Such a conclusion is also valid for Zeldovich matter 
($h=0$).  

Let us now investigate the behavior of the non-viscous 
anisotropic model at $t_s=0$. 
For Solution~II ($h\neq 0$), by using Eq.\,(5.10), we obtain in the 
main order $\exp[\gamma_0]\sim \exp[-At]$ when $t\rightarrow+\infty$ 
($t_s\rightarrow +0$), then $t_s\sim\exp[-At]$. By substituting the latter 
relation into  Eqs.\,(4.26), we obtain in the main 
order  
\begin{equation}\label{5.11} 
{\rm e}^{x^i}\sim 
|t_s|^{-\alpha^i/A+1/(D-1)}, \quad \rho\sim |t_s|^{h-2} \quad 
(t_s\rightarrow +0).  
\end{equation} 
According to Eq.\,(\ref{5.3}), the 
model for $h\in(0,2]$ has a Kasner-like behavior near the singularity (at 
$t_s=0$). Such a behavior describes the contraction of some spaces 
$M_1,\ldots,M_n$ and the expansion for the other ones. According to Eq.\,(\ref{5.2}), the
number of either contracting or expanding spaces depends on $n$ (the total number of spaces) and
$N_{i=1,n}$ (their dimensions), but there is at least a contracting manifold and expanding
one. Such dynamics is a mechanism of extra dimensions 
compactification (within the multidimensional cosmology).

One can easily show that the non-viscous anisotropic model for Zeldovich 
matter ($h=0$) described by Solution~IV has almost the same properties but 
with the second constraint given Eq.\,(\ref{5.2}) for the Kasner 
parameters substituted by 
$\sum_{i=1}^{n}N_i(\varepsilon^i)^2=\varepsilon$, where $\varepsilon$ is a 
constant such that $1/(D-1)<\varepsilon<1$.

\subsection{Models with bulk viscosity}
Let us now investigate the  models with bulk viscosity.
\subsubsection{}
The {\em viscous isotropic model\/} shows interesting features. The shear 
viscosity is not significant in this case and the model is described for 
$h\neq 0$ by  Solution II with $\alpha^i=0$ for $i=1,\ldots,n$. If 
$C_1C_2<0$ and $h>0$ then the solution can be written as follows~: for 
$i=1,\ldots,n$
\begin{eqnarray}
{\rm e}^{x^i}
&=&R_i\left(1- \exp\left[\frac{D-1}{D-2}\zeta_0(t-t_0)\right] \right)^{-2/[h(D-1)]},\label{5.12}\\
p_i^*
&=&\left(1- h\exp\left[-\frac{D-1}{D-2}\zeta_0(t-t_0)\right] 
\right)\rho,\label{5.13}\\ 
\rho
&=&\frac{D-1}{D-2}\frac{2\zeta_0^2}{\kappa^2h^2} \prod_{i=1}^n R_i^{-2N_i}
\exp\left[2\frac{D-1}{D-2}\zeta_0(t-t_0)\right]\nonumber\\
&&\times\left(1- \exp\left[\frac{D-1}{D-2}\zeta_0(t-t_0)\right] 
\right)^{2(2-h)/h]},\label{5.15}\\ s &=& 
\frac{D-1}{D-2}\frac{2B\zeta_0^2}{\kappa^2h^2} \prod_{i=1}^n R_i^{-hN_i} 
\exp\left[2\frac{D-1}{D-2}\zeta_0(t-t_0)\right] +s(-\infty),
\end{eqnarray}
where we use the set of independent constants $t_0,R_1,\ldots,R_n$, defined such that
\begin{equation}\label{5.17}
\frac{C_1}{C_2}=-\exp\left[\frac{D-1}{D-2}\zeta_0t_0\right],\quad
|C_1|^{-2/[h(D-1)]}\exp[\beta^i]=R_i.
\end{equation}
Let us consider this solution on the interval $(-\infty,t_0)$ for
$h\in(0,2)$. Then the synchronous time $t_s$ changes during the evolution 
on the interval $(-\infty,+\infty)$. Such a solution is non-singular and 
describes a monotonic isotropic expansion of the $D$-dimensional Universe 
with Friedman-like stage as $t_s\to +\infty$. 
The density $\rho$ increases from zero in the infinite past to some 
maximum value and then decreases to zero at the Friedman-like stage. The 
maximum of the density is reached at $t\equiv 
t_m=(D-2)\ln[h/2]/[\zeta_0(D-1)]+t_0$, when $p_i^*=-\rho$, see 
Eq.\,(\ref{5.13}). We have $p_i^*+\rho<0$ for $t\in(-\infty,t_m)$ and 
$p_i^*+\rho>0$ for $t\in(t_m,t_0)$, so the weak energy condition, see 
Eq.\,(\ref{5.6}), is not satisfied on the time interval $(-\infty,t_m)$.  
The entropy per baryon monotonically increases during the evolution and 
tends to some constant value in the infinite future.

The nonsingular solution obtained by Murphy [35] within flat FRW model 
with bulk viscosity for another second equation of state: $\zeta={\rm 
const}\rho$ exhibits the similar properties except the violating of the 
weak energy condition. The scale factor monotonically increases from zero 
in the infinite past and tends to the infinity at the Friedman stage of 
the evolution. The density monotonically decreases from some constant 
value in the infinite past and tends to zero according to Eq.\,(5.9) in 
the infinite future. The weak energy condition is valid for 
$t_s\in(-\infty,+\infty)$ (but the strong energy condition 
$\rho+3p^*\geq 0$ is not satisfied on the interval $(-\infty,t_s^*)$, 
where $t_s^*$ is some constant), so from this point of view the Murphy 
solution is more attractive.

\subsubsection  {}
 Now, let us study the properties of the {\em anisotropic viscous model\/} 
by taking into account only the bulk viscosity. Such model is described by 
Solution II with non-zero Kasner-like parameters $\alpha^i$ and has been 
previously studied \cite{GavrilovEtal95a,GavrilovEtal96a}. If $C_1C_2>0$ 
in Solution~II with $\alpha^i\neq 0$ then the density
has negative values at some stage of the evolution, 
which means that the weak energy condition is not satisfied. Hence, 
hereafter such solutions are not considered. If $C_1C_2<0$, then  Solution~II can be written as
follows~: for $i=1,\ldots,n$

\begin{eqnarray}
{\rm e}^{x^i}&=&
R_i \left|\sinh[\tilde{A}h(t-t_0)/2]\right| ^{-2/[h(D-1)]}
\exp\left[\left(\alpha^i-\frac{\zeta_0}{h(D-2)}\right)t\right],\label{5.18}\\
p^*_i
&=&\left(1-\frac{h\cosh^2\tilde{a}}{1+\sinh\tilde{a}
\sinh[\tilde{A}h(t-t_0)+\tilde{a}]}\right)\rho,\label{5.19}\\
\rho
&=&\frac{\langle \alpha,\alpha\rangle}{2\kappa^2}\prod_{i=1}^n R_i^{-2N_i}
\left|\sinh[\tilde{A}h(t-t_0)/2]\right|^{2(2-h)/h}
\exp[2\tilde{\zeta}_0t]\nonumber\\
&&\times\left(1+\sinh\tilde{a}\sinh[\tilde{A}h(t-t_0)+\tilde{a}]\right),\label{5.21}\\
\dot{s}
&=&\frac{B\zeta_0A^2}{\kappa^2}\prod_{i=1}^n R_i^{-hN_i}
\exp\left[\frac{D-1}{D-2}\zeta_0\left(1-\frac{2}{h}\right)t\right]
\sinh^2[\tilde{A}h(t-t_0)]\nonumber\\
&&\times\sinh^2[\tilde{A}h(t-t_0)+\tilde{a}],\label{5.22}
\end{eqnarray}
where we use independent constants $t_0,R_1,\ldots,R_n$ defined by
\begin{eqnarray}\label{5.23}
C_1=-\frac{C}{2}\exp[\tilde{A}ht_0/2],\quad
C_2=\frac{C}{2}\exp[-\tilde{A}ht_0/2],\\
R_i=|C|^{-2/[h(D-1)]}\exp[\beta^i],
\end{eqnarray}
and $\tilde{a}$ is defined by
\begin{equation}\label{5.24}
\sinh{\tilde{a}}=\frac{\tilde{\zeta}_0}{A},\quad
\cosh{\tilde{a}}=\frac{\tilde{A}}{A}.
\end{equation}
Let us note that for $t\in(-\infty,t_0)$ the density,  given by 
Eq.\,(5.19), has  negative values. In the following, this solution 
is only used on the interval $(t_0,+\infty)$.

If $h\in(0,2)$ then the harmonic time interval $t~:~(t_0,+\infty)$ corresponds to the following
synchronous time interval $t_s~:~(-\infty,t^0_s)$, where we choose the integration constant
$t^0_s=0$. We can easily prove that such a model has the stage of isotropic contraction by the
Friedman-like law defined in Eq.\,(\ref{5.9}) in the infinite past. Near the final point of the
evolution $t_s=0$ we obtain in the main order
\begin{equation}\label{5.25}
{\rm e}^{x^i}\sim t_s^{-\alpha^i/(\tilde{A}+\tilde{\zeta}_0)+1/(D-1)},
\ i=1,\ldots,n.
\end{equation}
The final point of the evolution $t_s=0$ is singular ($\rho\to +\infty$ as
$t_s\to -0$), and the characteristic of the singularity depends on the parameter
$\tilde{\zeta}_0/A$ (which determines the ratio between the viscosity parameter $\tilde{\zeta}_0$
and the anisotropy parameter $A$). If the ratio $\tilde{\zeta}_0/A\gg 1$ then we obtain from
Eq.\,(5.24) $\exp[x^i]\sim t_s^{1/(D-1)}$ near $t_s=0$.  In such a case, 
the model describes the contraction of all spaces $M_1,\ldots, M_n$ in the 
vicinity of the singularity. If $\tilde{\zeta}_0/A\ll 1$ then we obtain 
>from Eq.\,(5.24) $\exp[x^i]\sim t_s^{\varepsilon^i}$ (i.e. the 
singularity is of the Kasner type). According to Eq.\,(5.20), in 
both cases the model describes the unbounded production of the entropy at 
the final stage of the evolution.

Therefore, anisotropic Solution~II for $h\in(0,2)$ and $C_1C_2<0$ describes the model with
Friedman-like isotropic contraction in the infinite past and the anisotropic Kasner-like behavior
near the final singularity if the parameter $\tilde{\zeta}_0/A$ is small enough. Under such a
condition, the behavior of scale factors remains (qualitatively) the same as for the
anisotropic non-viscous model for $t_s<0$. Let us note that 
this model satisfies the dominant energy condition during the evolution. 
According to  Eq.\,(5.18), the ratio $p^*_i/\rho$ increases 
monotonically from the value $(1-h)$ at the Friedman-like stage in the 
infinite past, and tends to $1$ in the vicinity of the final singularity. One may also consider
Solution~II for $h>0$ and $C_1=0$, $C_2\neq 0$. This partial solution has the similar dynamical
behavior at the final stage of the evolution and satisfies the dominant energy condition as
$p_i^*=\rho$, $\forall t_s$.

 If the dominant energy condition is ignored then we may also consider 
 anisotropic Solution~II for $h<0$ and $C_1C_2<0$. In such a case the 
harmonic time interval $t~:~(t_0,+\infty)$ corresponds to the synchronous 
time interval $t_s~:~(0,+\infty)$ . We easily show that such a model has 
the Friedman-like singularity, defined in Eq.\,(\ref{5.9}) at 
$t_s=0$. The anisotropic behavior is possible far from the singularity, if 
the parameter $|\tilde{\zeta}_0|/A$ is small enough. This solution 
satisfies the weak energy condition as the ratio $p_i^*/\rho$ 
monotonically decreases during the evolution from the value $(1-h)$ at the 
Friedman-like stage, and tends to $1$ in the infinite future. Let us note 
that the asymptotical behavior of this solution far from the singularity 
($t_s\to +\infty$) is given by the Solution~II for $h<0$ and $C_2= 0$, 
$C_1\neq 0$. To investigate the possible anisotropic behavior far from the 
singularity let us consider this partial solution. It may be written in 
the synchronous time as following~: for $i=1,\ldots,n$ 

\begin{eqnarray}
{\rm e}^{x^i}
&=&R_it_s^{\alpha^i/(\tilde{A}+|\tilde{\zeta}_0|)+1/(D-1)},\quad
t_s>0,\label{5.26}\\
p^*_i=
\rho
&=&
\frac{D-2}{D-1}\frac{|\tilde{\zeta}_0|}
{\kappa^2(\tilde{A}+|\tilde{\zeta}_0|)}t_s^{-2},\label{5.29}\\
s&=&
\frac{(D-2)B|\zeta_0|(\tilde{A}+|\tilde{\zeta}_0|)}{(D-1)\kappa^2}
t_s^{-h}+s(0),\label{5.30}
\end{eqnarray}
where the constants $R_1,\ldots,R_n$ are such that
\begin{equation}\label{5.32}
\prod_{i=1}^n R_i^{N_i}=\tilde{A}+|\tilde{\zeta}_0|.
\end{equation}
It is evident from Eq.\,(5.25) that 
$\alpha^i/(\tilde{A}+|\tilde{\zeta}_0|)+1/(D-1)\to \varepsilon^i$ (Kasner 
parameter) as $|\tilde{\zeta}_0|/A\to 0$. Therefore, if the parameter 
 $|\tilde{\zeta}_0|/A$ is small enough, then the model describes the 
 contraction of a part of the spaces $M_1,\ldots,M_n$ and the expansion 
of another part. According to Eq.\,(5.27), the 
solution  describes the unbounded entropy production as $h<0$.

We note, that the anisotropic viscous model described by Solution~IV~
($h=0$) does not satisfy the weak energy condition as the density 
has negative values at some stage of the evolution.

\subsection{Models with bulk and shear viscosity}
Now let us investigate the model with both bulk and shear 
viscosity.  Such a model for $h\neq 0$ is described by Solution~III with 
non-zero Kasner like parameters $\alpha^i$ and $\nu>0$. By using the 
following properties of the modified Bessel functions \cite{Handbook} 
\begin{equation}\label{5.33}
I_{\nu+1}(\tau)<I_{\nu}(\tau),\quad
K_{\nu+1}(\tau)>K_{\nu}(\tau)\quad
\forall \tau\in(0,+\infty),\nu\geq 0,
\end{equation}
it can be proved that the density given  has negative 
values at some interval of time if $C_2=0$. If $C_2\neq 0$ then the 
Solution~III can be written as follows: for $i=1,\ldots,n$
\begin{eqnarray} {\rm e}^{x^i} &=&R_i 
\left(\tau^{-\nu}\left| CI_{\nu}(\tau)+K_{\nu}(\tau) \right|\right) 
^{-2/[h(D-1)]}\exp\left[\alpha^i{\rm e}^{-\eta_0 t}\right],\label{5.34}\\
\dot{x}^i
&=&\frac{2\eta_0\tau}{h}\left[\left(
\frac{CI_{\nu+1}(\tau)-K_{\nu+1}(\tau)}{CI_{\nu}(\tau)+K_{\nu}(\tau)}+1
\right)\frac{1}{D-1}-\varepsilon^i\right],\label{5.35}\\
p^*_i
&=&\left(1-h+\frac{h}{\tau} \frac {(D-1)\varepsilon^i-1-\nu
\frac{CI_{\nu+1}(\tau)-K_{\nu+1}(\tau)}{CI_{\nu}(\tau)+K_{\nu}(\tau)}} {\left(
\frac{CI_{\nu+1}(\tau)-K_{\nu+1}(\tau)}{CI_{\nu}(\tau)+K_{\nu}(\tau)}
\right)^2-1}\right)\rho,\label{5.36}\\
\rho
&=&\frac{2\eta_0^2}{h^2\kappa^2}\frac{D-2}{D-1}
\prod_{i=1}^n R_i^{-2N_i}\tau^{2(1-2\frac{\nu}{h})}
\left|CI_{\nu}(\tau)+K_{\nu}(\tau)\right|^{\frac{4}{h}}\nonumber\\
&&\times\left[\left( \frac{CI_{\nu+1}(\tau)-K_{\nu+1}(\tau)}{CI_{\nu}(\tau)+K_{\nu}(\tau)}
\right)^2-1 \right],\label{5.38}\\
\dot{s}
&=&\frac{4\eta_0^3B}{h^2\kappa^2}\frac{D-1}{D-2} \prod_{i=1}^n R_i^{-hN_i}
\tau^{2(1-\nu)}\nonumber\\
&&\times\left( \nu[CI_{\nu+1}(\tau)-K_{\nu+1}(\tau)]^2 +[CI_{\nu}(\tau)+K_{\nu}(\tau)]^2
\right),\label{5.39}
\end{eqnarray}
where the independent constants $R_1,\ldots,R_n,C=C_1/C_2$ are used. 
The variable $\tau$ was determined in Eq.\,(\ref{4.22}).
Also
we introduced Kasner parameters $\varepsilon^i$ by
\begin{equation}
\varepsilon^i={\rm sgn}[h]\frac{\alpha^i}{A}+\frac{1}{D-1}.
\end{equation}
By using the properties given in Eq.\,(4.29) and the asymptotical behavior 
of the modified Bessel functions \cite{Handbook},
one may prove that the density given in Eq.\,(5.33) has no negative 
values during the evolution only for~:  
\begin{enumerate} 
\item partial 
solution with $C=0$ on the interval $(-\infty,+\infty)$ of the harmonic 
time; 
\item 
solution with $C<0$ on the interval $(t_0(\tau_0),+\infty)$, 
where $\tau_0$ is the root of the equation 
\begin{equation}\label{5.43} 
CI_{\nu}(\tau_0)+K_{\nu}(\tau_0)=0.  
\end{equation} 
\end{enumerate} 
We note that the asymptotical behavior of the solution with $C<0$ as $t\to 
+\infty$ is given by the partial solution with $C=0$.

 Let us now investigate the models having a singularity at the beginning 
 of the evolution. Such solutions arise when $h<0$. In this case 
for the solutions with $C<0$ one obtain the following correspondences 
$(-\tau)\in(-\tau_0,0)\Leftrightarrow t\in(t_0(\tau_o),+\infty)
\Leftrightarrow t_s\in(0,+\infty)$. The singularity at  $t_s=0$
is of Friedman type. 
The shear viscosity leads to the isotropization at the
final stage of evolution. We obtain from Eq.\,(5.30-5.33) 
 $\exp[x^i]\sim t_s^{1/(D-1)}$ and $p_i^*=\rho\sim t_s^{-2}$ as $t_s\to 
+\infty$, i.e. the  model describes the isotropic Friedman-like expansion 
corresponding to Zeldovich matter ($h=0$). 
The anisotropic behavior is possible on some interval of time after the 
Friedman-like behavior. One can prove that if the constant $|C|\neq 0$ is 
small enough then on some interval the sign of the Hubble parameter 
$\dot{x}^i$ coincides with the sign of the Kasner parameter 
$\varepsilon^i$ (see Eq.\,(5.31) for $h<0$). However,
such regime is possible only on limited interval of time, because the 
final stage of the evolution exhibits the isotropic expansion due to the 
shear viscosity.

\begin{center}
Acknowledgments
\end{center}
This work was supported in part by the Russian State Committee for Science
and Technology, the Russian Fund of Basic Sciences and by Universit\'e de Provence (for
V.N.M.).
\pagebreak


\begin{thebibliography}{99}

\bibitem{Handbook} Abramowitz A, Stegun I A, 1964 {\it Handbook on 
Mathematical Functions\/} ( National Bureau of standards, Applied 
Math. Series 55).

\bibitem{BelinskyKhalatnikov75} 
Belinsky V A and Khalatnikov I M 1975 {\it Soviet Phys. JETP Letters\/} 
{\bf 21\/} 99.

\bibitem{BelinskyKhalatnikov76} 
Belinsky V A and Khalatnikov I M  1976 {\it Soviet Phys. JETP\/} {\bf 
42\/} 205.

\bibitem{BelinskyKhalatnikov77} 
Belinsky V A and Khalatnikov I M 1977 {\it Soviet Phys. JETP\/} {\bf 
45\/} 1.

\bibitem{BerezinEtal89}
Berezin V A, Domenech G, Levinas M L, Lousto C O and Umerez N D 1989 {\it 
Gen.  Relativ. Grav.\/} {\bf 21\/} 1177.

\bibitem{BleyerEtal91} 
Bleyer U, Liebscher D-E and Polnarev A G 1991 {\it Class. Quantum Grav.\/} 
{\bf 8\/} 477.

\bibitem{BleyerEtal94} 
Bleyer U, Ivashchuk V D, Melnikov V N and Zhuk A I 1994 {\it Nucl. 
Phys.\/} {\bf B429\/} 177.

\bibitem{CaderniFabbri78} 
Caderni N and Fabbri R 1978 {\it Nuovo Cimento\/} {\bf 44B\/} 228.

\bibitem{ChodosDetweiler80} 
Chodos A and Detweiler S 1980 {\it Phys. Rev.\/}  {\bf D21\/} 2167.

\bibitem{CollinsStewart71} 
Collins C B and Stewart J M 1971 {\it Monthly Notices Roy. Astron. Soc.\/} 
{\bf 151\/} 419.

\bibitem{DemianskyPolnarev90} 
Demiansky M and Polnarev A G 1990 {\it Phys. Rev.  D\/} {\bf 41\/} 3003.  


\bibitem{ForgacsEtal79} 
Forgacs P and Horvath Z 1979 {\it Gen. Relativ. Grav.\/} {\bf 11\/} 205. 


\bibitem{GleiserEtal85} 
Gleiser M, Rajpoot S and Teylor J G 1985 {\it Ann. Phys.\/} (NY) {\bf 
160\/} 299.

\bibitem{GavrilovEtal95} 
Gavrilov V R, Ivashchuk V D and Melnikov V N 1995 {\it J. Math. Phys.\/} 
{\bf 36\/} 5829.

\bibitem{GavrilovEtal95a} 
Gavrilov V R, Melnikov V N and Novello M 1995 {\it Gravitation and 
Cosmology\/} {\bf 1\/} 149.

\bibitem{GavrilovEtal96} Gavrilov V R, Ivashchuk V D and Melnikov V N, 
1996 {\it Class. Quantum Grav.\/} {\bf 13\/} 3039 .  

\bibitem{GavrilovEtal96a} 
Gavrilov V R, Melnikov V N and Novello M 1996 {\it Gravitation and 
Cosmology\/} {\bf 4(8)\/}.  

\bibitem{GavrilovEtal96b} 
Gavrilov V R, Melnikov V N and Triay R Exact solutions in multidimensional 
cosmology with bulk and shear viscosity {\it Preprint} CPT-96/P.3396, 
Marseille, France.  


\bibitem{Gron90} Gron O 1990 {\it 
Astrophys. Space Sci.\/} {\bf 173\/} 191.  

\bibitem{IvashchukMelnikov88} 
Ivashchuk V D and Melnikov V N 1988 {\it Nuovo 
Cimento\/} {\bf B102\/} 131.  

\bibitem{IvashchukMelnikov89} 
Ivashchuk V D and Melnikov V N 1989 {\it Phys.  Lett.\/} {\bf A135\/} 465. 
  
\bibitem{IvashchukEtal89} 
Ivashchuk V D, Melnikov V N and 
Zhuk A I 1989 {\it Nuovo Cimento B\/} {\bf 104\/} 575.  

\bibitem{Ivashchuk92} 
Ivashchuk V D 1992 {\it Phys. Lett.\/} {\bf A 170\/} 16.

\bibitem{IvashchukMelnikov94} 
Ivashchuk V D and Melnikov V N 1994 {\it Int. J. Mod. Phys.\/} {\bf D3\/} 
795.

\bibitem{IvashchukMelnikov95} 
Ivashchuk V D and Melnikov V N 1995 {\it Class. Quantum Grav.\/} {\bf 
12\/} 809.

\bibitem{Klimek76} Klimek Z 1976 {\it Nuovo Cimento\/} {\bf 35B\/} 249. 


\bibitem{KoikawaYoshimura85} 
Koikawa T and Yoshimura M 1985 {\it Phys. Lett. B\/} {\bf 155\/} 137. 


\bibitem{Lorenz84} Lorenz-Petzold D 1984 {\it Phys. Lett. B\/} {\bf 
149\/} 79.

\bibitem{Lorenz85} 
Lorenz-Petzold D 1985 {\it Phys. Lett. B\/} {\bf 158\/} 110.

\bibitem{Lukacs76} 
Lukacs B 1976 {\it Gen. Rel. Gravit.\/} {\bf 7\/} 635.

\bibitem{Melnikov94} Melnikov V N 1994 {\it Multidimensional Classical  
and Quantum Cosmology and Gravitation:  Exact Solutions and Variations of 
Constants\/} In: Cosmology and Gravitation. Ed. M.Novello. Editions 
Frontieres, Singapore, p. 147.

\bibitem{Melnikov95} Melnikov V N, {\it Multidimensional Cosmology and 
Gravitation\/}, CBPF-MO-002/95, Rio de Janeiro, Brasil.

\bibitem{Misner68}  
Misner C W 1968 {\it Astrophys. J.\/} {\bf 151\/} 431. 


\bibitem{MottaTomimura92} 
Motta D and Tomimura N 1992 {\it Astrophys. J.\/} {\bf 401\/} 437.

\bibitem{Murphy73} 
Murphy G 1973 {\it Phys. Rev.\/} {\bf D8\/} 4231.

\bibitem{NovelloAraujo88} 
Novello M and Araujo T A 1988 {\it Phys. Rev.\/} {\bf D22\/} 260.

\bibitem{OliveiraSalim88} 
Oliveira H P and Salim J M 1988 {\it Acta Phys. Pol.\/} {\bf B19\/} 649. 


\bibitem{Romero88} 
Romero C 1988 {\it Rev.Bras. de Fisica\/} {\bf 18\/} 75.

\bibitem{Sahdev84} 
Sahdev D 1984 {\it Phys. Rev. D\/} {\bf 30\/} 2495.

\bibitem{Stewart68} 
Stewart J M 1968 {\it Astrophys. Lett.\/} {\bf 2\/} 133.

\bibitem{Szydlowski88} 
Szydlowski M 1988 {\it Gen Relativ. Grav.\/} {\bf 20\/} 221.

\bibitem{SzydlowskiPajdosz89} 
Szydlowski M and Pajdosz G 1989 {\it Class. Quant. Grav.\/} {\bf 6\/} 
1391.

\bibitem{Tosa86} 
Tosa Y 1986 {\it Phys. Lett.\/} {\bf B174\/} 156.

\bibitem{Weinberg71} 
Weinberg S 1971 {\it Astrophys. J.\/} {\bf 168\/} 175.

\bibitem{WessonPonce94} 
Wesson P S and Ponce de Leon J 1994 {\it Gen. Rel.  Gravit.\/} {\bf 26\/}
555.

\bibitem{Wiltshire87} 
Wiltshire D L 1987 {\it Phys. Rev. D\/} {\bf 36\/} 1634.

\bibitem{Wolf89} 
Wolf C 1989 {\it Phys. Scripta\/} {\bf 40\/}, 9.

\bibitem{Zhuk95} 
Zhuk A I 1995 {\it Gravitation and Cosmology\/} {\bf 1\/} 119.

\end{thebibliography}
\end{document}